# On Space Debris Removal by Lasers: Can Spatially and Temporally Shaped Laser Pulses Be Advantageous for Propulsion?

Nadezhda M. Bulgakova

FZU - Institute of Physics of the Czech Academy of Sciences, Na Slovance 1999/2, 182 00 Prague 8, Czech Republic
nadezhda.bulgakova@fzu.cz

**Abstract**

For exploration of space, in particular in attempts to find new extra-terrestrial resources, human society has encountered the problem of space pollution with human-made debris, which represents high risks for space missions. This prompted extensive activities for cleaning the space using various techniques, which are briefly overviewed here. But the main focus of this paper is on using lasers for space debris removal. The attention is drawn to laser beam shaping techniques, which are discussed as potential technologies for deorbiting space debris, providing more energetically favorable laser propulsion compared to conventional laser beams.

## 1. Introduction

The space around the Earth is strongly polluted by debris, which is not only stones travelling in space and captured by Earth's gravity, but mostly human-made non-operating satellites and their fragments originating due to cosmic collisions. Although the problem of cleaning near-Earth space was raised shortly after starting active space exploration [1], nowadays it becomes critical. According to the report of the European Space Agency in 2024 [2], the estimated populations of objects in space are ~54.000 of size greater than 10 cm, ~1.2 million of size between 1 cm and 10 cm, and ~130 million with size from 1 mm to 1 cm. Even relatively small objects represent a real danger to space missions as, travelling with a velocity of 7-8 km/s, they can induce considerable surface degradation of spacecrafts or even their catastrophic disintegration [3,4]. In 2020[th], the launch traffic near the low Earth orbit (LEO) is strongly activated and the probability of catastrophic collisions due to accumulating debris grows exponentially [4]. Thus, accumulation of fragments of human-made satellites, if not apply specific measures for their removal, can make spaceflights in the future too risky, if not impossible. Additionally, it has been predicted [5,6] that the anthropogenic effect from uncontrolled re-entering of space debris to the Earth's atmosphere can become significant in comparison to the natural injection. Figure 1 presents an artistic view of the space debris problem [7].

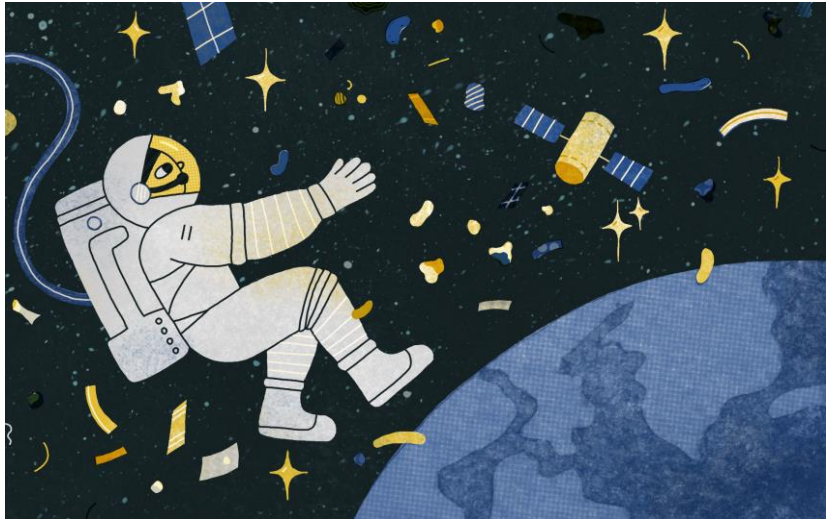

**Figure 1.** Artist's view on pollution of near-Earth space. Courtesy of B. Bulgakova [7].

Although the impacts on the atmosphere of the Earth, its climate, and ecosystem are not well studied yet, pollution of the stratosphere by metals resulting in aerosol formation, influence on the ozone layer and the D layer ion chemistry, and effects on the radiation balance create a danger for the Earth's population [6,8]. In the composition of the human-made debris orbiting the Earth, metals such as Al, Ti, Ni, Fe, Co, Au, and others dominate [3,5,9] with the major part (30-40%) of aluminum while solar array materials are also abundant where silicon still represents an essential part while Ge, GaAs, and Ga-InP become increasing contributors to space debris [5]. At present, approximately 2500 unfunctional satellites are orbiting around the Earth [10] with masses varied from 1 kg to several tons. According to evaluations made by Leonard and Williams [9], a scrap material in the orbits can be evaluated from ~5 to ~19 kilotons, thus, recycling such materials in the orbit would be very valuable for circular economy [9,11,12].

Thus, the tasks of the LEO cleaning from human-made debris, both by de-orbiting and in-orbit recycling, become vital for future exploration of space and hopefully, in the long perspective, for enriching the Earth by mineral resources from other planets.

There are many conceptual ideas and methods for space debris removal. Examples of comprehensive reviews can be found in [13-18]. Classifications of the methods are rather very wide and can vary in different literature sources. Shortly, they can be divided into contact and contactless ones. Contact methods include mostly using robotic arms, net capturing, tether, and harpoon mechanisms [13]. These methods are suitable for spacecrafts out of operation and relatively larger fragments, at least >10 cm. Some of these methods have already been verified under space conditions. Thus, the RemoveDEBRIS space mission has successfully demonstrated the abilities of the net and harpoon mechanisms [19]. Robotic manipulators, which are already essential parts in many space missions, are foreseen as perspective technology for cleaning space [20-23]. Among the proposed contactless methods, the most known are based on collimated ion beam thrust [24], electrostatic deflection of space objects by electron beams [25], laser propulsion [26-33], and using the gravity force [34]. The collimated ion beam (or ion beam shepherd) mechanism is based on directing a high-velocity ion flux on a space object, thus transferring momentum and changing its orbit. One of the disadvantages of this method is that the shepherd spacecraft experiences recoil pressure and, hence, needs a counteracting propulsion system to support a stable orbit. The electrostatic tractor concept is based on emitting an electron beam located on a serving satellite toward a debris object and changing it, thus providing an electrostatic force between the servicer and the object that can be used to pull the object to another orbit. The gravity tractor concept utilizes the idea of deflecting large, hazardous objects, such as asteroids, from possible collisions by using the mass of a spacecraft. This requires a long period for a small gravitational force action and can be appropriate for large objects.

These three methods are suitable for deorbiting relatively large objects. Among the contactless methods, lasers attract major attention since the laser ablative propulsion concept was introduced [35]. This method is considered to be most suitable for the removal of fragmentation debris with a size in range of 1–10 cm, which are the main hazards for colliding the large objects, leading to their fragmentation [2]. Although using lasers for space debris removal was sometimes criticized in the past, with the development of super-powerful lasers, which can send directed beams from ground, and relatively compact but powerful lasers, which can be positioned on a spacecraft, using laser propulsion for space debris removal becomes increasingly feasible [31]. It is important that, using an appropriate navigation system for tracking space debris, laser propulsion can efficiently be applied to deorbit relatively small space debris that represents a big hazard for space flights due to their tremendous and growing abundance in LEO [36,37].

By date, the real experiments on debris removal in space have not been performed. However, laboratory experiments performed in different countries have demonstrated the underlying concepts of using lasers and their feasibility. As examples, the European project CLEANSPACE [30], experimental campaigns in China [38], using a novel concept of fibre-based laser system [31], experimental missions on the technical feasibility of changing debris trajectories at DLR, Germany [39], NASA projects for mitigation of space debris risks [40], and international collaborations [41,42] can be listed. Laser-based concepts are grounded by experimental and theoretical proofs that the ablation of material with a ground- or space-based laser can create enough thrust for deorbiting or deflecting debris. However, much research should be done for optimization the momentum transfer regarding laser power, wavelength, material properties of the targeted debris.

In this Opinion paper, the laser propulsion method for space debris removal is discussed from the perspective of novel optics techniques, which potentially can make this method more efficient than envisioned nowadays. This concerns laser beam shaping techniques, both spatial and temporal shaping. In section 2, a short historical overview of laser propulsion is given, with first ideas to use this technique for cleaning LEO. Section 3 is focused on three potential techniques of laser beam shaping, which can be applied to make space debris removal more efficient, and an outline of necessary research topics is given. In section 4, a short summary is presented. The paper does not claim the completeness of citation of the relevant publications but only crops papers that influenced the author's opinion.

## 2. Laser Propulsion as the Technique for Space Debris Removal

The concept of laser propulsion was well formulated in the historical paper by Kantrowitz [35], where he proposed the idea of using ground-based lasers for ablation-induced propulsion of spacecrafts to the Earth's atmosphere. Already in the next years, this paper provoked a burst of activities, both theoretical and experimental, in which researchers suggested different schemes of laser propulsion as well as different types of lasers, cw and pulsed, that can be used for ablative acceleration of various rockets [43-52]. Nowadays, this field is further bursting with numerous reviews and new ideas on how to optimize the thrust [53-57]. The application of laser propulsion for cleaning space becomes vital, as discussed above.

To our knowledge, the idea of using space-based lasers was proposed in 1991 [58], and after that, it was experimentally investigated for dielectric and metallic targets by Kuznetsov and Yarygin [59]. To measure the recoil pressure generated by laser interaction with the sample, Kuznetsov [60] used a piezoelectric transducer attached to the rear (non-irradiated) surface of the sample. The piezoelectric transducer converts the recoil pressure into an electrical signal, which is then analyzed and can be converted to the pressure values; the latter can be used for the evaluation of momentum transfer. Additionally, the experimental scheme used in [60] enabled measurements of the transmission of laser light through the ablation plume, which allowed for evaluating the laser energy balance (see discussion below). This scheme was used in a series of works where the pressure pulsations on the targets irradiated by laser pulses of ~800 µs duration were measured at different ambient pressures [61,62]. In these works, the modeling based on the Navier-Stokes equations enabled to conclude that pressure pulsations were developed due to the involvement of the ambient gas in motion. Another mechanism, which can lead to pressure pulsations on target ablated with relatively long laser pulses, can be periodic laser breakdown of ablation products followed by screening the target by developed plasma, its expansion, a new portion of ablation, and so on, as was proposed by Rykalin et al. [63]. This effect, which looks to be highly possible for relatively long, powerful laser pulses or cw lasers suitable for space propulsion, is not yet considered in detail.

In this century, for the topic of laser removal of space debris for cleaning LEO to make space missions safer and minimize risks of collisions, various strategies have been proposed (see [29-33] and references therein). For this, the measurements of laser energy coupling with different materials present in space debris with different laser parameters are of high importance. A sophisticated method for momentum transfer measurements is based on a solid pendulum on which a target for laser action is mounted [42]. Being irradiated by a laser, the pendulum starts to oscillate, and its velocity is usually measured using Doppler velocimetry or an interferometry method as discussed in [42]. The experiments were performed with 400-fs and 80-ps laser pulses at a 1057-nm wavelength for irradiating Al, Ta, W, Au, and polyoxymethylene targets, and the measurement method showed an accuracy of ±10%. The authors of [42] provided an extensive survey of the literature on optimal fluences $F_{opt}$ vs. pulse duration (reproduced in Figure 2) for laser momentum transfer to large variety of materials (metals, polymers, graphite, grafoil, carbon phenolic and fiber reinforced polymer composites). For the designations employed in Figure 2, the readers are referred to the original paper [42].

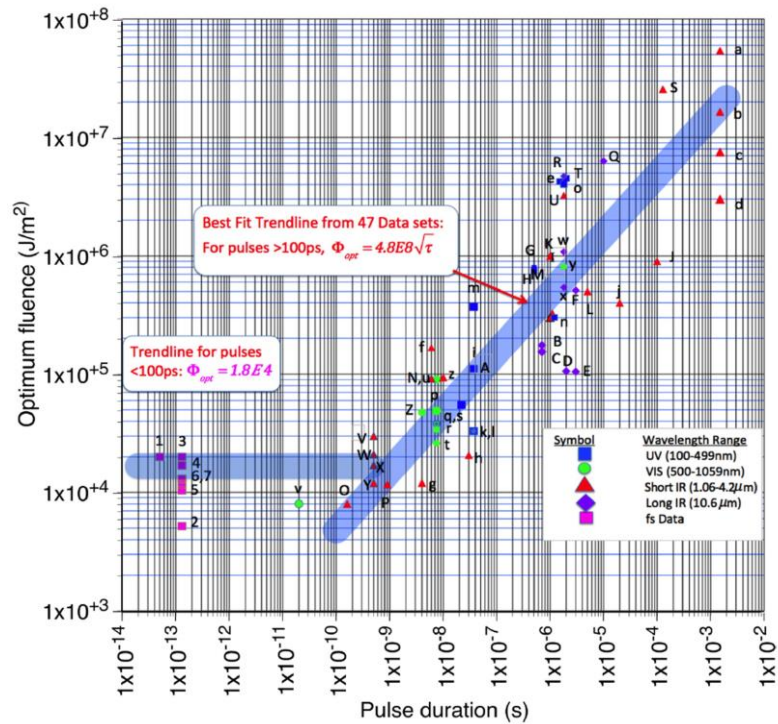

**Figure 2.** Literature values for optimum fluence across a wide range of pulse durations (see designations in the original paper [42]). On the right (pulses longer than 100 ps), the trend is for $F_{opt}$ to increase with the square root of pulse duration. Reprinted from [42], Laser impulse coupling measurements at 400 fs and 80 ps using the LULI facility at 1057 nm wavelength. J. Appl. Phys. 2017, 122, 193103, Copyright 2017, with permission of AIP Publishing.

The observed increase of $F_{opt}$ as a square root of pulse duration indicates that femto- and picosecond laser pulses are more energetically favorable for momentum transfer than longer ones. Indeed, for ultra-short pulses, heat diffusion is negligible during the laser pulse [64]. As a result, the absorbed energy in the surface layer of the irradiated material is consumed to localized heating, melting, and ablation if the ablation threshold is exceeded. The main mechanism of ablation at ultrashort laser pulses is phase explosion (or explosive boiling), which is the disintegration of matter to atoms and clusters/droplets [65]. One drawback for space debris removal can be in creation of more debris, though of very small size, hence less dangerous for space flights. For longer laser pulses, the laser-affected region is defined by material heat diffusion, which is usually much larger than the laser light absorption depth. If considering laser energy balance, the ratio between the energy spent for ablation (positive energy from the viewpoint of laser propulsion) and the energy loss for target heating (negative energy) is decreasing with pulse duration, which is especially pronounced for materials with high thermal conductivity, as well demonstrated in Figure 2.

In general, in laser irradiation of different materials for inducing their ablation with the development of high propulsion thrust, the laser energy balance is one of the key factors. At ultrashort laser pulses, material ablation starts after the laser pulse action, which conditions the absence of target screening by ablation products. The situation is very different for nanosecond and longer pulses, which was already mentioned above for the case of pressure pulsations on laser irradiated targets with 800 μs pulses. An example of laser energy balance for nanosecond laser ablation of a metal (niobium) and a semiconductor (silicon which is abundant in LEO as fragments of old-fashioned solar cells [5]) is given in Figure 2. The simulations were performed based on the thermal model, taking into account laser energy shielding by the plasma emerging above the target during the laser pulse action [66,67]. Modeling has been performed from the ablation threshold until the fluences at which the transition from thermal vaporization to phase explosion occurs and, hence, the thermal model is not applicable anymore [65,67]. The irradiation conditions were as follows: 1064-nm wavelength, 13-ns FWHM pulse duration for niobium (experimental conditions of Ref. [67,68]) and 266 nm, 3 ns FWHM (experimental conditions of Ref. [69]). As is evident from Figure 2, such fluences of rather gentle ablation cannot be very efficient for inducing propulsion, as plasma shielding is very pronounced. According to simulations for silicon, the thermal (kinetic) energy of the vaporizing atoms, which conditions the recoil pressure generation, is ~4% starting from laser fluences higher than 9 J/cm$^2$. For niobium, the situation is even worse: the thermal energy of atoms leaving the surface during ablation does not exceed 0.3%. With the transition to the phase explosion mechanism, the laser-generated thrust should strongly increase due to much more massive ablation [67,69] with emission of the vapor-droplet mixture that should, however, be investigated from the viewpoint of laser propulsion.

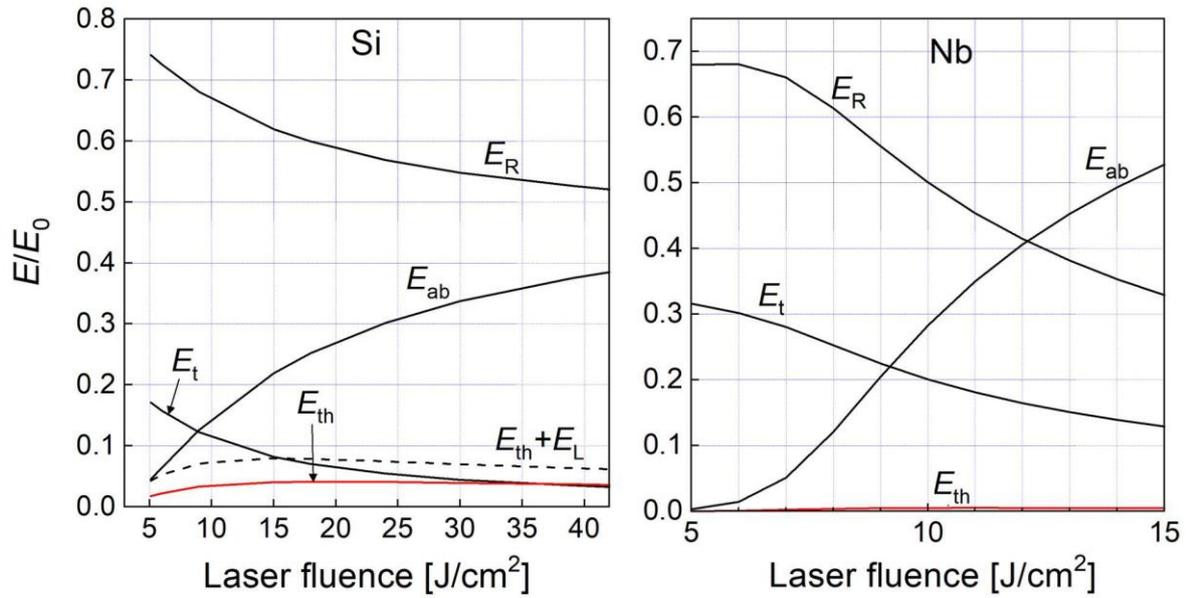

**Figure 3.** Distribution of the laser energy density between different channels for laser ablation of silicon and niobium as a function of laser fluence. $E_0$ is the incident laser energy; $E_t$ are the losses to the target due to heat conduction; $E_R$ is the reflected energy; $E_{ab}$ is the energy absorbed in the plume; $E_L$ is the part corresponding to the latent heat of vaporization; $E_{th}$ is the thermal energy of the vaporized particles.

Considering the role of plasma shielding upon laser irradiation of materials, a comment should be made on plasma feedback, which can enhance ablation, while still investigated rather poorly. For the case of graphite, it has been shown [70] that expanding plasma plume emits radiation in the UV spectral range, a part of which is efficiently absorbed by the ablated target. This provides deeper absorption and stronger ejection of material from the laser-irradiated zone. This mechanism may compensate for some losses of laser energy due to energy re-emission toward the target and hence enhance the propulsion efficiency. However, this topic requires further studies in terms of kinds of material, laser wavelength and fluence, and pulse duration.

Soulard et al. [31] presented a rigorous analysis of requirements for lasers to utilize them in the removal of space debris with a size of 10 cm. For ground-based lasers, they evaluated that 10-ns, 1-µm laser pulses with energy of 400 J can be successfully focused to a distance of ~800 km for providing $E_{opt}$ on the debris surface of ~5 J/cm$^2$ that corresponds to the optimal value of the coupling efficiency for the case of aluminum debris. The evaluation of focusing 100-ps beams of space-based lasers with achieving $E_{opt} \sim 1$ J/cm$^2$ gives ~100 km distance of focusing at a pulse energy of 77 J. We note that such fluences are sufficient for the ablation of any kind of material.

Considering the price of lasers, it looks that, for space propulsion, lasers with pulse durations in the range from ~100 ps to ~10 ns could be most suitable for applications in space-based debris deorbiting. However, such laser pulses have their pros and cons. From the point of view of their delivery to orbit, they are lightweight as compared with ultrafast lasers, while the optimal fluence for propulsion is not much higher. However, to increase their power for providing efficient propulsion in space, additional components are needed that can make such lasers heavier. Other means, which are not extensively investigated, concern laser beam shaping techniques that can provide more efficient laser energy coupling, as will be discussed in the next section.

## 3. Laser Beam Shaping for Optimization of Momentum Transfer

Laser beam shaping, both spatial and temporal, has shown strong advantages for material processing. This includes spatial beam shaping for generation of flat-top (or top-hat) and doughnut-shaped beams with or without angular momentum [71-78], application of double pulses separated in time when one pulse pre-excites the material while the next pulse induces more efficient ablation than a single pulse with the same total energy [79], a variation of double pulse technique with two pulses at different wavelength [80-83], introducing diffractive optical elements which split laser beam into several or numerous beams that is especially important for high power lasers and enables high-speed material micromachining [84-87], and finally efficient ablation that can be achieved with burst-mode lasers [88-90]. Schematically, these techniques, which can be potentially used for optimization of space debris removal, are summarized in Figure 4. Also, adaptive optimization [91-93] should be mentioned, which can enable adapting the laser beam shape for achieving efficient ablation of specific materials. Below, possible implications for the mentioned types of shaping regarding laser propulsion are considered.

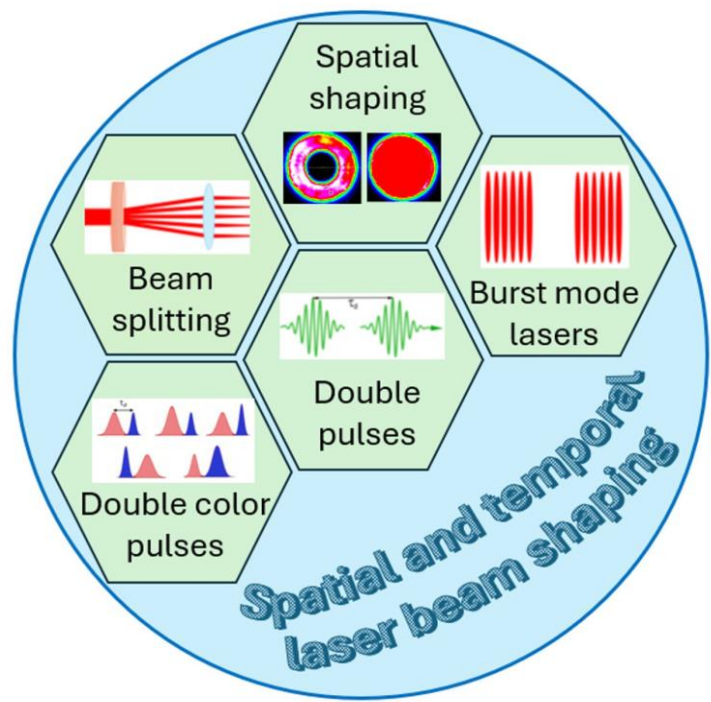

**Figure 4.** Schematics of techniques of laser beam shaping, which can be used for optimization of space debris removal.

*3.1. Adaptive Optimization of Laser Pulses*

Temporal shapes of short and ultrashort laser pulses can play an important role in material processing via determining the strength and interplay of physical processes depending on the temporal variation of the intensity input into the matter and thus enabling to direct the processes in a predefined route. Adaptive optimization of the laser waveform is based on an iterative procedure that controls a certain process parameter via a feedback loop until the desired optimal result has been achieved [91-93]. This technique, which does not require prior knowledge of the detailed physics and/or chemistry of the process, can be applied to any kind of material independently of the structure, thermophysical, and optical properties. One can speculate that tailoring the laser beam shapes via adaptive optimization can be an efficient methodology for adjusting the laser parameters in order to enhance ablation rates of random debris encountered on the way of spacecrafts.

Figure 5 presents the simulation results of laser irradiation of silicon under the irradiation conditions of Ref. [91], where an adaptive optimization technique was used to achieve enhancing of the ablated ion velocity, and hence this case is relevant to the propulsion issue. Here, the modeling was repeated using the refined model described in [94]. The black line corresponds to the lattice temperature evolution with time for the optimal laser pulse action at 800-nm wavelength as found in [91], which consists of a 170-fs pre-pulse followed by an 8-ps pulse. The total fluence was 0.8 J/cm$^2$, with a quarter of it coupling to the target in the pre-pulse. For a single femtosecond laser pulse, the model included the mechanism of ultrafast melting as the fluence exceeded its threshold [90]. A simplified consideration was as follows. Starting from the time moment when 10% of valence band electrons have been excited to the conduction band, the absorption and reflection coefficients are linearly increasing with time during 400 fs (ultrafast melting is typically observed in several hundred femtoseconds after irradiation [95]) to reach the value of molten silicon while the electron and hole mobilities are correspondingly decreasing. The result is shown in Figure 5 by the blue line. The results obtained with the refined model differ only slightly from those reported in [91]. Thus, the main conclusion is that the optimal pulse leads to efficient heating of material well above the thermodynamic critical point (~7500 K [96]) while the femtosecond pulse alone heats silicon well below the critical point. Thus, such a difference in laser-induced heating of the surface layer should correspondingly result in strong differences in the velocities of the ejected material that is observed in [91]. Considering a significant increase in ion velocity and ion yield (~3 times) observed in the experiments for an optimized laser pulse, one can expect the momentum transfer enhancement up to an order of magnitude as compared to a single Gaussian pulse that requires experimental verification.

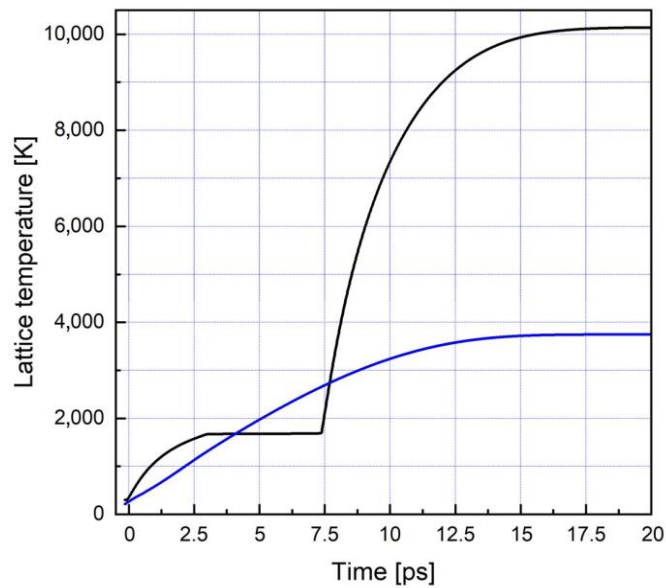

**Figure 5.** Calculated temperature temporal profile on the surface of a silicon sample irradiated by the adaptively optimized laser pulse with 170-fs pre-pulse followed by 8-ps pulse (black line) and the single 170-fs pulse (blue line). The total fluence in both pulses was the same, 0.8 J/cm$^2$.

It should be mentioned that the temporal shaping schemes are flexible and can be optimized to only a few iterations [97], which makes them practical in different technologies, including space applications.

*3.2. Double Pulses with Variable Separation Time*

Double laser pulses with time separation between them applied to material ablation, which are considered as a part of temporal laser beam shaping, demonstrate a higher efficiency compared to Gaussian beams under certain conditions [79] that depend on time separation. The effect of a strong reduction of the ablated crater depth in the regimes of ultrashort double pulse irradiation of metals was discovered by Semerok and Dutouquet [99]. This effect was explained by the authors as plasma shielding of the second laser pulse with corresponding heating of the plasma plume generated by the first laser pulse, which was observed as increasing plasma intensity. Later,

Povarnitsyn et al. [100] performed hydrodynamic simulations and found the generation of the shock wave in the expanding plasma plume, which suppresses the material ablation by the second pulse. Certainly, both mechanisms can be present while both can serve for the optimization of laser propulsion. Indeed, if the ablated material is reheated by the second laser beam, a part of it can be redeposited with bringing its momentum back to the target. Additionally, the shock wave generated in the plasma plume will also bring momentum to the target. For the regime simulated in [100], the shock wave generated by the second laser pulse has a pressure amplitude of ~2 GPa, leading to substantial redeposition of material ablated by the first pulse. Roughly, this can result in at least a doubling of the momentum transfer as compared to a single laser pulse of the same energy as in two subsequent pulses. Thus, double ultrashort laser pulses can prove to be highly advantageous for space ablative propulsion and space debris removal, which calls for further studies.

*3.3. Dual Wavelength Laser Pulses*

The variation of double pulse laser ablation is the application of two pulses with different wavelengths, where the first pulse excites the material efficiently while the second pulse efficiently heats the excited electrons, and hence matter, which is more pronounced for bandgap materials [83]. Using bi-chromatic laser pulses has proven to enhance the efficiency of laser energy coupling [101,102] and materials ablation [81,82]. Experimentally, it has been shown that the ablated mass can be increased by 2-3 times as compared with a single laser pulse of the same energy at a wavelength that produces stronger ablation [82,83]. It should be noted that the ablation efficiency depends on the sequence of laser pulses of different wavelengths and the time separation between them, which calls for further studies. However, the momentum transfer should be proportional to the ablated mass and can be optimized via providing an efficient ablation. Another advantage is that dual laser pulses can be beneficial for processing of hard materials [80] that can be met in space. Thus, the dual-wavelength regimes may be good candidates for space debris removal, although the momentum transfer measurements have not been performed yet.

*3.4. Burst Mode Lasers*

In recent years, burst mode lasers have been increasingly utilized for the ablation of different materials, mostly in the laser microfabrication field. It has been found that this technique, which uses bursts of ultrafast laser pulses instead of single pulses, enhances laser ablation efficiency as well as the surface quality [88-90]. This approach allows for the control of laser energy coupling into materials, resulting in higher ablation precision. Žemaitis et al. [90] have demonstrated the increase in the material removal efficiency by 18.0 %, 44.5 %, and 37.0 % for Al, Cu, and stainless steel when using bursts instead of single pulses. Thus, potentially the burst-mode technique may be applicable for more efficient, at least by dozens of percent, momentum transfer in space propulsion and space debris removal areas.

*3.5. Spatial Beam Shaping*

Detailed review on spatial and temporal beam shaping and its applications can be found in Ref. [98]. The application of top-hat laser beams has advantages in pulsed laser deposition and surface processing. As an example, such beams allow highly reproducible laser-induced forward transfer of thin films for printing light-emitting diode pixels [73]. The application of top-hat beams enables better control of both width and depth of the ablated area as compared to Gaussian beams [76]. Top-hat laser beams produced flatter/shallower weld pools compared to Gaussian beams in laser-powder bed fusion [78]. Doughnut-shaped (DS), radially and azimuthally polarized laser beams demonstrated advantages for deterministic formation of micro-needles that may have helicity in the case of non-zero orbital momentum and for the generation of surface periodic patterns [72,74,75,77]. It can be imagined that, due to specific dynamics of the ablation plume generated by a DS laser pulse with redeposition of ablated material back to the target, the momentum transfer to the target can be more efficient as compared to Gaussian laser beams. The effects connected with the redeposition of the ablated material will be shortly discussed in the next sub-section.

It should be noted that DS laser pulses with the same laser energy and the same laser spot diameter as for a Gaussian laser pulse have a peak fluence 2.7 times smaller [103] that should be considered at focusing. The authors have shown that the DS pulses behave at focusing similarly to the Gaussian ones in a diffraction-limited manner. Thus, the same optical system can be used for both Gaussian and DS beams, including the beam expanders to radii necessary for ground-based or space-based lasers, from ~1 m to dozens of meters, depending on laser wavelength and focusing distance.

For the ground-based lasers of both spatial shapes, atmospheric turbulence should have a similar effect (beam spreading, fluctuations of intensity, and its decrease) due to random changes of refractive index and scattering on aerosols along the path of laser beam propagation [104], which puts the space-based lasers into an advantageous position.

*3.6. Beam Splitting Using Diffractive Optical Elements*

With the fabrication of diffractive optical elements (DOE) enabling the splitting of a powerful laser beam into numerous sub-beams, the material processing field undergoes substantial progress due to a highly increased throughput of surface micromachining [84-86]. For the sake of space debris deorbiting, this technique could be extremely useful not only due to the possibility to utilize very powerful laser pulses without the risk of decay of debris pieces into numerous fragments but also because laser ablation plumes upon their intercollisions may create considerable back flows, thus bringing additional momentum back to the irradiated target [105,106]. However, the existing DOEs represent diffractive beam splitters which, being designed for specific wavelengths, are characterized by a separation angle between output beams [107]. The separation between beams is drastically increasing with the propagation distance. Thus, the existing DOE may be suitable for manipulation by big debris. A solution to suppress diffraction would be beneficial for many applications on Earth and in space.

**4. Conclusions**

In this overview of the laser ablation techniques related to space debris removal, the field that becomes vital for space exploration and communication technologies, an attempt is made to attract the attention of researchers to potentially new possibilities that are obtained due to the development of novel optical means for laser beam manipulation. This relates to spatial and temporal beam shaping, which, under certain conditions, may strongly enhance the efficiency of laser ablation and, hence, increase momentum transfer to the targeted space objects. Such techniques could provide opportunities to use laser energy in more optimal ways. To achieve this, a large body of research is required for investigations of laser momentum transfer to different types of materials present in space. The issue should also be mentioned concerning the shape and size of the harmful objects floating in the LEO space. Pulsed laser interaction with objects of irregular shapes is still challenging, although it represents an important problem for efficient cleaning of the near-Earth space [108,109] from let small debris, which however, are most abundant in space.

Finally, it is necessary to mention the risks of using high-power lasers for space debris removal. The potential risks include fragmentation of debris into smaller pieces with creation more debris which still can be hazardous, damaging operational satellites or leading to their orbital changes, back-reflected laser light which can damage sensitive equipment [110,111]. Thus, it is necessary to develop strong mitigation strategies and scenarios for preventing such risks that should include methodologies for precise laser targeting in order to avoid operational satellites and implementing robust tracking and control systems for monitoring laser interactions in space. From this viewpoint, space-based laser systems with high accuracy of focusing look as a most promising step for cleaner and safer space.

**Acknowledgments**

The research was partially funded by the European Regional Development Fund and the State Budget of the Czech Republic (Project SENDISO No. CZ.02.01.01/00/22_008/0004596). The author thanks Dr. Claude R. Phipps, Dr. Yoann Levy, Prof. Alexander V. Bulgakov, and Dr. Martin Divoky for stimulating and helpful discussions, which resulted in the preparation of this paper.